\documentclass{article}

\usepackage{amsmath}  % load first
\usepackage[english]{babel}
\usepackage[utf8]{inputenc}
\usepackage{amssymb}
\usepackage{graphicx}
\usepackage{authblk}
\usepackage{caption}
\usepackage{multirow}
\usepackage{amscd}
\usepackage{geometry}
\usepackage{a4wide}
\usepackage[usenames]{color}
\usepackage{etaremune} 
\usepackage{hyperref}
\usepackage{amsfonts}
\usepackage{ifthen}
\usepackage{color}
\usepackage{amscd}
\usepackage{latexsym}

\newcommand{\bd}{\begin{definition}}
	\newcommand{\ed}{\end{definition}}
\newcommand{\bt}{\begin{theorem}}
	\newcommand{\et}{\end{theorem}}
\newcommand{\bi}{\begin{itemize}}
	\newcommand{\ei}{\end{itemize}}
\newcommand{\ben}{\begin{enumerate}}
	\newcommand{\een}{\end{enumerate}}
\newcommand{\beq}{\begin{equation}}
\newcommand{\eeq}{\end{equation}}

\newcommand{\R}{\mbox{$ \mathbb{R}  $}}
\newcommand{\C}{\mbox{$ \mathbb{C} $}}
\newcommand{\N}{\mbox{$ \mathbb{N} $}}

\newcommand{\pp}{\mbox{$ {\cal P} $}}

\newtheorem{definition}{Def.}[section]
\newtheorem{theorem}{Theorem}[section]

\def \qed{ \hfill $\Box$ \\}

\def \qed{ \hfill $\Box$ \\}

\begin{document}

\title{Geometry of color perception. Part 1:  Structures and metrics of a homogeneous color space}

\author[1]{Edoardo Provenzi\thanks{edoardo.provenzi@math.u-bordeaux.fr}}
%\author[1]{Author B\thanks{B.B@university.edu}}
%\author[1]{Author C\thanks{C.C@university.edu}}
%\author[2]{Author D\thanks{D.D@university.edu}}
%\author[2]{Author E\thanks{E.E@university.edu}}
\affil[1]{Université de Bordeaux, CNRS, Bordeaux INP, IMB, UMR 5251\\ F-33400, 351 Cours de la Libération, Talence, France}
%\affil[2]{}

\renewcommand\Authands{ and }
\date{}

\maketitle

%\begin{history}
%\received{(Day Month Year)}
%\revised{(Day Month Year)}
%\end{history}

\begin{abstract}
This is the first half of a two-part paper dealing with the geometry of color perception. Here we analyze in detail the seminal 1974 work by H.L. Resnikoff, who showed that there are only two possible geometric structures and Riemannian metrics on the perceived color space $\pp$ compatible with the set of Schrödinger's axioms completed with the hypothesis of homogeneity. We recast Resnikoff's model into a more modern colorimetric setting, provide a much simpler proof of the main result of the original paper and motivate the need of psychophysical experiments to confute or confirm the linearity of background transformations, which act transitively on $\pp$. Finally, we show that the Riemannian metrics singled out by Resnikoff through an axiom on invariance under background transformations are not compatibles with the crispening effect, thus motivating the need of further research about perceptual color metrics.
\end{abstract}

%\keywords{Schr\"{o}dinger's axioms of color space; Stiles' measure; Homogeneous spaces; Lie groups and algebras. MSC codes: 57S25, 57S99, 78-02}

\section{Introduction and state of the art}

This first half of a two-part paper provides a thorough review and a critical analysis of the pioneering work of H.L. Resnikoff on color perception developed within the papers \cite{Resnikoff:74,Resnikoff:74bis,Resnikoff:75} and the book \cite{Resnikoff:book}. These works are amongst the major inspirations for a modern program of re-foundation of colorimetry that will be discussed in the second part, in which it will be shown how to recast Resnikoff's model in a quantum-like theory via the framework of Jordan algebras.

Even if it may be surprising at first glance, Resnikoff belongs to a vast ensemble of mathematicians, physicists, biologists, philosophers and even poets fascinated by the concept of color. The list is impressive, ranging from Plato to Wittgenstein, passing through Aristotle, Descartes, Hook, Newton, Euler, Young, Helmholtz, Maxwell, Grassmann, Riemann, Goethe, Schopenhauer, Locke, Weber, Fechner, Dalton, Hering and, last but not least, Schr\"odinger, see e.g. \cite{Niall:2017} for a modern translation of Schr\"odinger's work on color.

In fact, it is the research about the mathematical analogies between optics and color, on one side, and the oscillating behavior of quantum particles, on the other, that led Schr\"odinger to propose the famous equation which bears his name in quantum mechanics \cite{Schroedinger:03}. As it will be recalled in section \ref{sec:Schroedinger}, Schr\"odinger performed a synthesis of the most important findings about the mathematical theory of color perception in a coherent set of axioms, introducing one of his own. This can be thought as a psycho-physical counterpart of what Maxwell did for electromagnetism. 

The experiments of Wright and Guild, see e.g. \cite{Wyszecky:82}, could have paved the way for a further development in the mathematical understanding of color perception, however, the then recently founded Commission Int\'ernational de l'\'Eclairage (CIE), took a more practical path by building up geometrically flat color spaces which had the advantage to be easier to handle for engineering purposes. While the XYZ space still stands today as a handy color space for colorimetry, its developments until recent years, see e.g. \cite{Schanda:07}, lacked of mathematical rigor and introduced ad-hoc parameters to adapt newly discovered phenomena to the existing color space structures, instead of modifying their geometry to fit the new observations.

Resnikoff's papers and book, instead, put in question the flat geometry of the color space. They were written in the middle of the '70s of the twentieth century, about the same period when researchers in relativistic quantum field theory developed the \textit{standard model} of fundamental physical interactions and when some first attempts to fuse quantum mechanics and general relativity into a single theory have been proposed. This \textit{zeitgeist} could explain the reason why Resnikoff decided to use techniques which are quite common in theoretical physics (as Riemannian geometry, homogeneous spaces, Lie groups and algebras) to study color perception. In this sense, his achievements could be considered a very elegant example of `theoretical psycho-physics'. 

In spite of its extreme originality and deepness, Resnikoff's work remained practically unnoticed until today, probably due to the fact that the mathematical knowledge needed to understand the meaning of his findings is quite vast and does not belong to the typical mathematical background of scientists working on colorimetry.

One of the aims of this paper is to rewrite Resnikoff's results in more modern and pedagogical terms, thus making them accessible to a wider range of researchers in colorimetry, vision science and image processing. 

We will discuss in particular detail the homogeneity axiom for the space of perceived colors $\pp$, which led Resnikoff to show that, if we accept the homogeneity hypothesis, $\pp$ can only be the canonical Helmoltz-Stiles space well-known to colorimetrists, or a completely new space of hyperbolic nature. We will provide an alternative and much simpler proof of this result, underlining also an error in the original demonstration proposed by Resnikoff in \cite{Resnikoff:74}. 

Differently from the standard colorimetric setting, Resnikoff established his theory in what we call today \textit{color in context}, i.e. a color stimulus embedded in a uniform background. This aspect reveals to be crucial for the development of his theory, because the group of transformations acting transitively on $\pp$ are identified with background changes. However, we remark that a major issue remains open: background transformations must be linear to fit in Resnikoff's theory, but no experimental data is available yet to confirm or confute this hypothesis. For this reason, we will discuss a psycho-physical experiment that can be used to verify the linearity of background transformations. 

Finally, we will show that the crispening effect contradicts Resnikoff's hypothesis that background changes are isometries with respect to the perceptual color metric. Thus underlying the need of further research about perceptual color metrics.

\section{Review of Schr\"odinger's axioms for the space of perceived colors}\label{sec:Schroedinger}
Schrödinger's axioms for the space of perceived colors can be resumed with the following statement: \textit{the space of perceived colors is a regular convex cone embedded in a real vector space of dimension less or equal to 3}. 

The validity of this sentence, however, is \textit{bounded within the limits of the standard observational conditions of colorimetry}, which, as we are going to discuss, are very restrictive and far from those of a natural visual scene. 

We start with the notation and nomenclature that we need for the rest of the paper. 
\medskip
\bi 
\item $\Lambda=[\lambda_\text{min},\lambda_\text{max}]$ denotes the \textit{visual spectrum}, the extrema of $\Lambda$ are left unspecified because their numerical values are not important and because there is no agreement about them. Typically, one chooses $\lambda_\text{min}=380$nm (extreme violet) and $\lambda_\text{max}$=780nm (extreme red).
\medskip
\item $\textbf{x}:\Lambda\to \R^+$: is the light function representing the electromagnetic radiation associated to the \textit{color stimulus}, or \textit{physical light}. A \textit{spectral color stimulus} $\textbf{x}_{\lambda_0}(\lambda)$ relative to the wavelength $\lambda_0$ is a narrow-band visible radiation, generally modeled by a Gaussian centered in $\lambda_0$ and with a small standard deviation (typically of the order of 1nm), or a piece-wise constant function everywhere null in $\Lambda$ apart from a small interval of wavelengths containing $\lambda_0$.

%for which there exist two values $k, \varepsilon >0$ such that $\textbf{x}_{\lambda_0}(\lambda)$ can be expressed as follows
%\beq
%\textbf{x}(\lambda)= \begin{cases}
%	k & \text{if } {\lambda_0} -\varepsilon < \lambda < {\lambda_0} +\varepsilon \\
%	0 & \text{otherwise.} 
%\end{cases}  
%\eeq 
\medskip
\item Since light stimuli have finite energy, i.e. $\int_\Lambda \textbf{x}(\lambda)^2 \, d\lambda < +\infty$, a physical light can be considered as an element of $L^2(\Lambda)\subset L^1(\Lambda)$, where the inclusion holds because the Lebesgue measure of $\Lambda$ is finite. Light stimuli are real-valued, so $L^2(\Lambda)$ will be considered as a real vector space and we will write
\beq 
L_+^2(\Lambda)=\{x:\Lambda \to [0,+\infty), \; x\in L^2(\Lambda)\},
\eeq
for the space of color stimuli.
\medskip
\item \textit{Standard colorimetric observational conditions}: in the conventional tests, a physical light \textbf{x} is presented in a \textit{dim room} to an observer with an aperture angle of either 2 degrees (foveal vision), or 10 degrees (extra-foveal vision). As we will see in section \ref{sec:Phomogeneous}, Resnikoff does not consider this standard observational conditions, since, in his model, \textbf{x} is presented as a small central area seen against a uniformly illuminated background. In this latter case we talk about color in (uniform) context. Experiments about color in non-uniform context are still quite rare, see e.g. \cite{Rudd:04}, \cite{Gronchi:17}, and confined to very simple geometric configuration, far from the complexity of natural scenes. In this section, only the standard colorimetric observational conditions will be considered.
\medskip
\item \textit{Color matching}: the typical way to compare the perception of two physical lights $\textbf{x},\textbf{y} \in L_+^2(\Lambda)$ is to divide the field of view in two (creating a \textit{bipartite field}) and putting the two color stimuli side-by-side. $\textbf{x}$ and $\textbf{y}$ are said to \textit{match} if no edge is perceived between them. We stress that this is not the only way to perceptually compare color stimuli, but it is by far the more common for the standard colorimetric observational conditions described above.
%\begin{figure}[h!]
%	\caption{\csentence{Bipartite field.}
%		Two color stimuli presented to an observer bipartite field of vision. The context (in blue in the picture) can be present or not.}\label{fig:bipartite}
%\end{figure}
\medskip
\item For any $\textbf{x},\textbf{y} \in L_+^2(\Lambda)$, we write $\textbf{x}\sim \textbf{y}$ to mean that $\textbf{x}$ and $\textbf{y}$ are perceived as identical\footnote{Resnikoff avoids using the term \textit{metameric equivalence}, which refers to the case when $\textbf{x}$ and $\textbf{y}$ have different spectral radiant power distribution, but they generate the same tristimulus values, \cite{Wyszecky:82}, i.e. the weights of three fixed primary colors that are needed to match a reference color. Following Resnikoff, we do not employ the metameric equivalence because the primaries do not intervene in his model.} in a color matching experiment. We stress that, when we write the expression $\textbf{x}\sim \textbf{y}$, the color stimulus $\textbf{x}$ is shown \textit{on one side} of the bipartite field and $\textbf{y}$ \textit{on the other side}. The symbol is well posed, since $\sim$ is indeed an equivalence relation \cite{Dubois:09}.
\medskip
\item The \textit{space of perceived colors} in the standard colorimetric observational condition is defined as the set of equivalence classes of equally perceived lights, i.e. the quotient space:
\beq\label{eq:pp}
\pp= L_+^2(\Lambda)/\sim \ . 
\eeq
The equivalence class of $\textbf{x}$ will be simply denoted by $x$: 
\begin{equation}
x=[\textbf{x}]_\sim \ .
\end{equation}
Given $x,y\in \pp$, $x=y$ means $[\textbf{x}]_\sim = [\textbf{y}]_\sim$, i.e. $\textbf{x} \sim \textbf{y}$, so that \textit{equality in $\pp$ means perceptual match of color stimuli}. The 0 of $\pp$ is the equivalent class of all physical lights that are so dim to be perceived as black. 
\medskip
\item We can endow $\pp$ with the operations of \textit{sum} and \textit{multiplication by a positive real scalar}. For any $\textbf{x},\textbf{y} \in L_+^2(\Lambda)$ and any positive scalar $\lambda\in \R^+$, $\textbf{x}+\textbf{y}$ is interpreted as superposition of light beams and $\lambda \textbf{x}$ as the intensity modulation of $\textbf{x}$ by the factor $\lambda$. These operations defined on $L^2_+(\Lambda)$ can be passed to the quotient space $\pp$ simply by defining:
\beq\label{eq:linpp}
\lambda_1 x+\lambda_2 y=[\lambda_1 \textbf{x}+\lambda_2 \textbf{y}]_\sim \ , \qquad \forall \lambda_1,\lambda_2 \in \R^+, \; \forall \textbf{x},\textbf{y} \in L_+^2(\Lambda),
\eeq
so $z=[\textbf{z}]_\sim$ can be written as $z=\lambda_1 x + \lambda_2 y$ if and only if  $\textbf{z}\sim \lambda_1 \textbf{x}+\lambda_2 \textbf{y}$. E. Dubois has proven in \cite{Dubois:09} that, if we consider the standard observational conditions of colorimetry, these operations are well-defined, in the sense that they do not depend on the choice of the representative in the perceptual equivalence class. To resume, in $\pp$ it is possible to operate \textit{conical} combinations of perceived colors, i.e. linear combinations with positive real coefficients. In particular, convex combinations of elements of $\pp$, i.e. conical combinations with coefficients summing to 1, are well-defined.

\medskip 
\item The smallest vector space containing $\cal P$ is: 
\beq 
V=\mathcal P - \mathcal P=\{x-y, \, \, x,y\in \mathcal P\},
\eeq 
see e.g. \cite{Faraut:94}. The elements of $V$ written as $-y$, with $y\in \cal P$, will be called \textit{virtual colors}. To understand the role of virtual colors in colorimetry, it is worthwhile recalling the famous Wright and Guild experiments, see e.g. \cite{Wyszecky:82} or \cite{Schanda:07}, which have shown that, for a fixed color stimulus $z\in L^2_+(\Lambda)$, there are three \textit{spectral} color stimuli $\textbf{x},\textbf{y},\textbf{w}\in L^2_+(\Lambda)$, and three real positive coefficients  $a,b,c\in \R^+$, such that either $a\textbf{x}+b\textbf{y}+c\textbf{w}\sim \textbf{z}$ or $a\textbf{x}+b\textbf{y}\sim \textbf{z}+c\textbf{w}$. In this last case, by recalling the convention above, we must superpose $\textbf{z}$ with $c\textbf{w}$ on one side of a bipartite field to color match the superposition $a\textbf{x}+b\textbf{y}$ on the other side.
This is where virtual colors enter into play: given $x,y,w,z\in \pp$, $a,b\in \R^+$ and $c<0$, $z=ax+by+cw$ belongs to $V$ but not to $\cal P$, because $c$ is negative. The colorimetric interpretation is the following: the color stimulus $a\textbf{x}+b\textbf{y}$ shown to an observer on one side of a bipartite field, matches $\textbf{z}+(-c)\textbf{w}$ shown on the other side, i.e. $a\textbf{x}+b\textbf{y}\in [\textbf{z}+(-c)\textbf{w}]_\sim$.
\ei 

\medskip

\noindent With this notation, Schr\"odinger's axioms, see \cite{Schroedinger:20}, can be stated like this. 

\bi
\item \textbf{Axiom 1} (Newton 1704, \cite{Newton:52}) If $x\in \pp$ and $\alpha \in \R^+$, then $\alpha x \in \pp$; \medskip
\item \textbf{Axiom 2} (Schr\"odinger 1920, \cite{Schroedinger:20}) if $x \in \pp$, $x\neq 0$,  then it does not exist any $y \in \pp$, $y\neq 0$, such that $x+y=0$;
\medskip
\item \textbf{Axiom 3} (Grassmann 1853, \cite{Grassmann:1853} \& Helmholtz 1866 \cite{Helmholtz:05}) For every
$x,y \in \pp$ and for every $\alpha \in [0,1]$, $\alpha x+(1-\alpha)y \in \pp;$
\medskip
\item \textbf{Axiom 4} (Grassmann 1853, \cite{Grassmann:1853}) For all quadruple of perceived lights $x_k\in \pp$, $k=1,\ldots,4$, there are coefficients $\alpha_k\in \R$, \textit{not all simultaneously null}, such that $\sum\limits_{k=1}^{4} \alpha_k x_k =0$.
\ei

Let us now discuss the colorimetric and mathematical meaning of the axioms. A finer version of Axiom 4 will be obtained mixing Axiom 1,2 and 4, furthermore, an important property of $\pp$ will be underlined as a consequence of Axioms 1 and 3.

Mathematically speaking, the meaning of Axiom 1 is simple: \textit{$\pp$ is an infinite cone} embedded in $V$. However, notice that this is an \textit{idealization}: when $\alpha$ gets very large, photoreceptors saturate until the glare limit is reached and we lose sight abilities. Instead, as $\alpha$ gets small we pass to the mesopic or to the scotopic conditions, in which both cones and rods or only the rods, respectively, are activated. If $\alpha$ approaches 0, then we lose our ability to see. Thus, the cone $\cal P$ is truncated both from above and below, with the shift from the photopic to the scotopic  condition (the so-called Purkinje effect \cite{Wyszecky:82}) still representing a major challenge from both a mathematical and a colorimetric point of view. The discussion of these important issues deserves a paper by its own and here we will just consider the idealized model of $\pp$ as an infinite cone.

Axiom 2 means that no superposition of perceived colors different from 0 is perceived as the absence of light\footnote{This assumption is \textit{true for non-coherent light} because, for coherent light, destructive interference can extinguish light intensity in certain spatial positions when two light beams are superposed.}. Mathematically speaking, this implies that the cone $\pp$ is \textit{regular} (also said \textit{proper}).

Axiom 3 means that the line segment which joins the perceived colors $x$ and $y$ consists entirely of perceived colors, thus Axioms 1,2 and 3 altogether imply that \textit{$\pp$ is a regular convex cone}. Axiom 3 is well known to be equivalent to be closed under conical convex combinations, i.e. linear combinations with positive coefficients between 0 and 1 whose sum is 1. 

This fact, along with Axiom 1, implies that $\pp$ is actually \textit{closed} under conical combinations, in fact, for all $\alpha_1,\alpha_2\in \R^+$ and $x_1,x_2\in \pp$, $\frac{1}{\alpha_1+\alpha_2}\left(\alpha_1x_1+\alpha_2x_2\right)\equiv z$ is a convex combination of elements of $\pp$, so $z\in \pp$ thanks to Axiom 3, but then also $(\alpha_1+\alpha_2)z=\alpha_1x_1+\alpha_2x_2\in \pp$ thanks to Axiom 1. By iterating this argument we have that $\sum\limits_{k=1}^{n} \alpha_k x_k \in \pp$ $\forall \alpha_k\in \R^+$, $x_k\in \pp$, $k=1,\ldots,n$. 

Axioms 1 and 3 imply that \textit{$\pp$ is a connected and contractible cone}, i.e. $id_{\cal P}$ is homotopic to a constant map \cite{Munkres:2000}. Intuitively, this means that $\pp$ can be continuously contracted to a single point. This simple remark will turn out to be crucial for the analysis of $\pp$.

Axiom 4 means that every collection of more than three perceived colors is a linearly dependent family in $V$.

A finer version of Axiom 4 can be obtained with the following argument. First of all, notice that Axioms 1-3 prevent the $\alpha_k$'s to have all the same sign. In fact, let us imagine that all the coefficients $\alpha_1,\ldots,\alpha_4$ are positive, then $\bar x = \sum\limits_{k=1}^3 \alpha_k x_k \in \pp$ (thanks to what just proven) and $\bar y = \alpha_4 x_4 \in \pp$ (thanks to Axiom 1), then $\sum\limits_{k=1}^{4} \alpha_k x_k =0$ implies $\bar x + \bar y = 0$, which is prevented by Axiom 2. A similar argument can be used when all the coefficients $\alpha_1,\ldots,\alpha_4$ are negative.

%If all the $\alpha_k$'s are negative, then:
%$$\sum\limits_{k=1}^{4} \alpha_k x_k = \sum\limits_{k=1}^3 \alpha_k x_k + \alpha_4 x_k = -\sum\limits_{k=1}^3 (-\alpha_k) x_k + \alpha_4 x_k \equiv -\sum\limits_{k=1}^3 \beta_k x_k + \alpha_4 x_4,$$
%$\beta_k >0$ $k=1,2,3$ and $\alpha_4 <0$, so $\sum\limits_{k=1}^3 \beta_k x_k \equiv \bar x \in \pp$ thanks to Axiom 1 and 3, and the relationship $\sum\limits_{k=1}^{4} \alpha_k x_k =0$ is equivalent to $-\bar x=-\alpha_4 x_4$, with $-\bar x\in \pp$ and $-\alpha_4 x_4=\bar y\in \pp$, but then we would have $-\bar x=\bar y$, $\bar x,\bar y\in \pp$, which is prevented by Axiom 2.

To resume, Axiom 1-4 imply this stronger version of Axiom 4.

\medskip
\noindent \textbf{Axiom 4'} \, For all quadruple of perceived lights $x_k\in \pp$, $k=1,\ldots,4$, there are coefficients $\alpha_k\in \R$, \textit{not all with the same sign}, such that $\sum\limits_{k=1}^{4} \alpha_k x_k =0$. 
\medskip

There are only two options coherent with Axiom 4'. The first option is that three coefficients have the same sign and the remaining one has opposite sign. Since equality in $V$ is color matching and a negative coefficient means that the corresponding light stimulus must be shown on the other side of the bipartite field, this means that one light stimulus color matches the superposition of other three light stimuli. 

In the second option, two coefficients are positives and two are negatives: this means that the superposition of two lights stimuli color match the superposition of other two lights stimuli. We thus see that Schrödinger's axioms are coherent with the experiments of Wright and Guild \cite{Schanda:07}.

Another direct consequence of Axiom 4 is that the dimension of $V$ is either 1,2 or 3. In particular, we call an observer for which:

\bi 
\item dim$(V)=3$: \textit{trichromate};
\item dim$(V)=2$: \textit{dichromate};
\item dim$(V)=1$: \textit{monochromate};
\item dim$(V)=0$: \textit{blind}. \\
\ei 

Following \cite{Ashtekar:99}, we observe that the \textit{projection map}
\begin{equation}
\begin{array}{cccl}
\pi : & L_+^2(\Lambda) & \longrightarrow & \pp  \\
& \textbf{x} & \longmapsto         & x,
\end{array}
\end{equation}
implies that \textit{infinitely many spectrally different lights coincide perceptually}. 

In what follows, we will fix our attention only on the trichromatic case, i.e. dim$(V)=3$, so that, from now on, \textit{$\pp$ will be interpreted as a regular convex cone embedded in a three dimensional real vector space}. 

\section{Resnikoff's homogeneity axiom for $\pp$}
As stated in the introduction, in \cite{Resnikoff:74} Resnikoff used the theory of homogeneous spaces to study the geometry and the metrics of the perceived color space $\pp$. This is far from being a trivial task, since the equivalence classes that make up $\pp$ are very difficult to characterize from a mathematical point of view. Thus, a theory of $\pp$ which bypasses the use of these equivalent classes is highly desirable. 

Understanding why Resnikoff considered homogeneity a paramount property in the analysis of $\pp$ is a key point. For this reason, it is worth recapping the basic information about homogeneous spaces.

If $X$ is a non empty set and $G$ is a group, then a map $\eta: G\times X \to X$, $(x,g)\mapsto g(x)$ is said to be a left action of $G$ on $X$ if $e(x)=x$, $e$ being the neutral element of $G$, and if $(gh)(x)=g(h(x))$ for all $g,h\in G$ and $x\in X$. In this case, $X$ is called a $G$-space. If we fix any $g$, then the map $\eta_g:X\to X$, $x\mapsto g(x)$ is bijective, its inverse being $\eta_{g^{-1}}$, because $g(x) \mapsto g^{-1}(g(x))=e(x)=x$. So, a left $G$ action on $X$ can equivalently be defined as a group homomorphism $G\to \text{Aut}(X)$, where Aut$(X)$ is the group of all automorphims (one-to-one functions) on $X$. Any element $g\in G$ is called a \textit{symmetry} for $X$.

Since $\pp$ is not merely a set, but a convex cone embedded in a vector space, we are more interested in $G$-spaces with an intrinsic structure. In this case, we call $X$ a $G$-space if the elements of $G$ preserve the structure of $X$, i.e. if Aut$(X)$ is the group of the automorphims of the category in which $X$ belongs, e.g. homeomorphisms for topological spaces, diffeomorphisms for smooth manifolds, invertible linear maps for vector spaces, and so on.

Fixed an arbitrary $x\in X$, the set $G\cdot x=\{g(x) \in X \; : \; g\in G\}$ is called the $G$-orbit of $x$. $X$ is said to be a \textit{$G$-homogeneous space} if $G\cdot x=X$ for every $x\in X$, i.e. if the action of $G$ on $X$ is transitive. In this case, for every couple of elements $x,y \in X$, there exists a transformation $g\in G$ such that $g(x)=y$, which explains why the concept of homogeneity is said to translate into mathematical terms the fact that \textit{no point of the space is special}. If $Y\subset X$, then we define the group Aut$(Y)=\{\varphi\in \text{Aut}(X) \; : \; \varphi(Y)=Y\}$ of automorphisms of $X$ that, restricted to $Y$ become automorphisms of $Y$, then $Y$ is \textit{$G$-homogeneous} if the action defined by the group homomorphims $G\to \text{Aut}(Y)$ is transitive on $Y$. 

Let us apply this to the case of interest for us, i.e. that of a general cone $\cal C$ embedded in a  real vector space $V$ of finite dimension $n$. In this case, if we define GL$(\mathcal C)=\{T\in \text{GL}(V), \; T(\mathcal C)=\mathcal C\}$, then $\cal C$ is said to be a \textit{homogeneous cone in} $V$ if, for any two points $a,b\in \cal C$, it exists $T\in $ GL$(\mathcal C)$ such that $b=T(a)$. We will also need a localized version of this property: $\cal C$ is {\em locally homogeneous cone in  $V$} if for every $a\in C$ there is an open neighborhood $U_a$ of $a$ such that every $b\in U_a$ it exists $T\in $ GL$(\mathcal C)$ such that $b=T(a)$, where open is referred to the Euclidean topology of $\cal C$ inherited by $V$.

\subsection{The 1-dimensional motivation to study $\pp$ as a homogeneous space}

Resnikoff declares that the motivation to study $\cal P$ as a homogeneous space comes from the analysis of Weber-Fechner's law \cite{Wyszecky:82} in metric terms. Weber-Fechner's law, often described as the first psycho-physical law ever determined, describes the perceptual response of humans with respect to changes of \textit{achromatic stimuli}, i.e. visual inputs that depend only on their intensity (typically obtained by activating only the retinal rods with dim lights).  Experiments showed that the perceptual counterpart of an achromatic stimulus of intensity $x\in \R^+=(0,+\infty)$, called \textit{brightness} and denoted with $b(x)$, is proportional to $\log x$ (for a wide range of intensities), thus, the relative brightness $b(x_1)-b(x_2)$ between two visual stimuli of intensity $x_1$ and $x_2$ is proportional to $\log(x_1)-\log(x_2)=\log\frac{x_1}{x_2}=\log \frac{\lambda x_1}{\lambda x_2}$ for all positive coefficient $\lambda$ belonging to the range of values for which Weber-Fechner's law is valid. This explains why the relative brightness is invariant under the simultaneous modification of light intensity expressed by
\beq  x_1 \mapsto \lambda x_1, \quad x_2 \mapsto \lambda x_2, \qquad \lambda >0. \eeq
$\R^+=(0,+\infty)$, interpreted as the set of all possible visible light intensities, is both a cone embedded in the real 1-dimensional vector space $\R$ and a group with respect to the ordinary multiplication of positive real numbers. The very simple observation that \beq \forall x,y\in \R^+, \; y=\frac{y}{x}x\equiv \lambda x,\eeq shows that \emph{$\R^+$ is a $\R^+$-homogeneous cone}. Weber-Fechner's law implies that the relative brightness between two perceived lights is a $\R^+$-invariant function defined on $\R^+$. What is crucial here is that, up to a selection of unit of measurement, the brightness difference expressed by Weber-Fechner's law coincides with the \textit{unique}  $\R^+$-invariant Riemannian distance\footnote{That is to say, the only Riemannian distance for which the multiplication by a positive scalar is an isometry, see section \ref{sec:Riemannmetrics} for more details about this.} on $\R^+$,
%that is induced by a Riemannian metric on the tangent space $T_1\R^+\cong \R$
i.e. \beq d(x_1,x_2)=|\log(x_1)-\log(x_2)|=\left|\log\frac{x_1}{x_2}\right|, \qquad x_1,x_2\in \R^+. \eeq
This consideration represented a major inspiration for Resnikoff, who extended these ideas to the 3-dimensional color space $\pp$.

\subsection{$\pp$ as a homogeneous space}\label{sec:Phomogeneous}
\begin{figure}[!ht]
	\centering
	\includegraphics[width=1.8in]{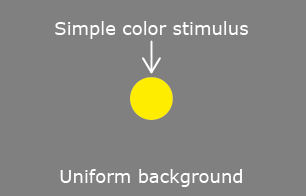}
	\caption{In Resnikoff's model a color is always associated to a couple given by a physical light and a uniform background illumination in which it is embedded.}\label{fig:context}
\end{figure}

Resnikoff's model for a homogeneous perceived color space is intimately connected with the non-standard observational configuration that he assumed, which is depicted in Figure \ref{fig:context}.

In this setting, a color is always associated to a light stimulus embedded in a uniform context. It follows that, in this context, the definition of $\pp$ given in (\ref{eq:pp}) must be modified as follows:  
\beq\label{eq:ppincontext}
\pp = \left( L^2_+(\Lambda) \times L^2_+(\Lambda) \right) \diagup_\sim \ ,
\eeq
i.e. a perceived color $x\in \pp$ here is defined as a perceptual equivalence class of couples $(\textbf{x},\textbf{b})\in L^2_+(\Lambda) \times L^2_+(\Lambda)$, where $\textbf{x}$ is the central light stimulus and $\textbf{b}$ is the light uniformly distributed on the background. Two distinct couples $(\textbf{x}_1,\textbf{b}_1)$ and $(\textbf{x}_2,\textbf{b}_2)$, $\textbf{x}_1\neq \textbf{x}_2$, $\textbf{b}_1\neq \textbf{b}_2$, belong to the same equivalence class if, for $i=1,2$, the central light stimuli $\textbf{x}_i$  embedded in the corresponding background $\textbf{b}_i$ induce the same perceived color $x\in \pp$. Resnikoff does not comment if, in this non-standard observational configuration, $\pp$ conserves the properties of a regular convex cone, this is a paramount important issue that will be discussed in section \ref{sec:linearityissue}. For the moment, we will assume that $\pp$ is a regular convex cone also in this setting.

As in section \ref{sec:Schroedinger}, we define the smallest vector space containing $\cal P$ as $V=\mathcal P-\mathcal P$. The only difference is that, in Resnikoff's setting, color match will be done for color stimuli embedded in a uniform background and not for stand-alone physical lights. This requires a novel specification of color matching that Resnikoff did not discussed.

Once established the model framework, we can begin with the mathematical construction that leads to the homogeneity axiom. First of all, since $\pp$ is a (positive) cone embedded in a three dimensional real vector space $V$, Aut$(\pp)$ will be given by  \textit{orientation-preserving linear transformations of $V$ which preserve $\pp$}, i.e.
\beq \text{GL}^+(\pp):=\{B\in \text{GL}(3,\R), \;\text{det}(B)>0 \text{ and } B(x)\in \pp \; \; \forall x\in \pp\}, \eeq
where GL$(3,\R)$ is identified with the group of invertible real $3\times 3$ matrices with determinant different from zero, i.e. the complementary set in M$(3,\R) \cong \R^9$ of det$^{-1}\{0\}$, the inverse-image of 0 by the determinant function, which is continuous in the Euclidean topology, thus det$^{-1}\{0\}$ is closed and so GL$(3,\R)$ is an open subset of $\R^9$. The request of positive determinant is introduced to respect the direction of each generatrix of the cone $\pp$. 

\begin{figure}[!ht]
	\centering
	\includegraphics[width=1.8in]{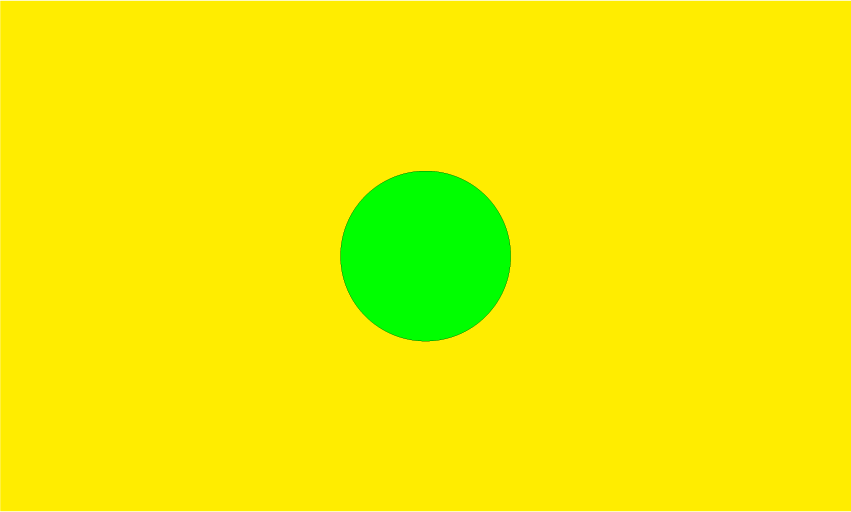}\;
	\includegraphics[width=1.8in]{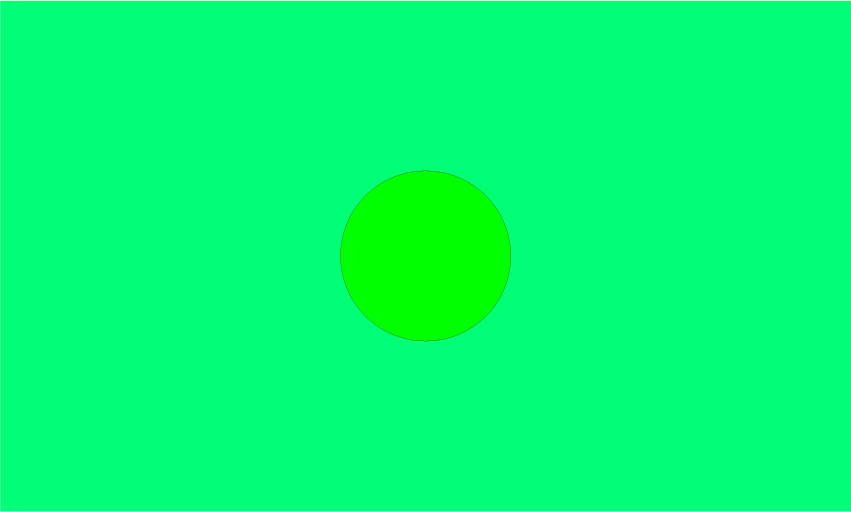}
	\caption{The inner disks appearing in the center of the two images are exactly the same physical color stimulus. However, the one in the left image is perceived, with respect to its background illumination, as a very saturated green, instead, after the change of background illumination shown by the image on the right, the color stimulus is perceived as yellowish. Due to the small size of the color stimuli on this document, they are surrounded by a thin black circle to enhance their visibility.}\label{fig:backgroundchange}
\end{figure}

Resnikoff claimed that, if we interpret $B\in$ GL$^+(\pp)$ as a `\textit{change of background illumination}', or \textit{background transformations} for short, then the action of GL$^+(\pp)$ on $\pp$ is transitive, thus making $\pp$ a homogeneous cone. The argument that he used follows this line of reasoning. First of all, it is generally accepted that any perceived color $x\in \pp$ can be transformed into any `sufficiently near' one $y\in \pp$ by an appropriate change of background illumination, see Figure \ref{fig:backgroundchange} for a graphical representation of this phenomenon.

For this reason, $\pp$ can be considered as a local homogeneous space with respect to the group GL$^+(\pp)$. Notice that this is \textit{not a physical property} of color, but a \textit{perceptual feature} of human vision, usually referred to as \textit{chromatic induction}, see e.g. \cite{Wallach:48,Rudd:04,Gronchi:17} for more details about how induction can be measured.

We now need some topological considerations. $\pp$ inherits the structure of metric space from $V$, thought as a three dimensional Euclidean space. Local homogeneity implies that, for every $x\in \pp$, there exists an open neighborhood $U_x \subset \pp$ such that each $y\in U_x$ can be expressed as $y=B(x)\in \pp$ for some $B\in$ GL$^+(\pp)$, so every element of $\pp$ is an interior point. To resume: \textit{$\pp$ is open in $V$}.

%the structure of a differentiable manifold from $V$\footnote{$V$, as a real vector space, is naturally endowed with the Euclidean differential structure.}. With respect to this differential structure, each transformation $B\in $ GL$^+(\pp)$, $B:\pp \to \pp$, is a diffeomorphism.

Let us now consider local homogeneity in conjunction with Axiom 3, i.e. with the convexity of $\pp$: for every couple of perceived colors $x,y\in \pp$, the line segment $L$ that joins $x$ to $y$ lies entirely in $\pp$. Local homogeneity assures that, for any $z\in L$, there exists an open neighborhood $U_z\subset \pp$ that is a homogeneous space with respect to the group GL$^+(\pp)$. As $z$ varies in $L$, we obtain the open covering $\bigcup\limits_{z\in L} U_z$ of $L$, and, since $L$ is a compact subset of $\pp$, we can extract a \textit{finite} open covering of $L$ from it, i.e. there exist $x_1,\ldots,x_n \in L$, $n<+\infty$, such that $\bigcup\limits_{k=1}^n U_{x_k}$ is an open covering of $L$. 

Let $B_k\in $ GL$^+(\pp)$ be the change of background illumination which carries $x_k$ to $x_{k+1}$, where $k=1,\ldots, n-1$, $x_0\equiv x$ and $x_n\equiv y$, then, since GL$^+(\pp)$ is a group, the transformation $B\equiv B_n\circ B_{n-1} \cdots \circ B_1$ carries $x$ to $y$, i.e. $y=B(x)$, for every couple of perceived colors $x,y\in \pp$. As a consequence, $\pp$ is a GL$^+(\pp)$-homogeneous space.

One might object that, operationally speaking, transforming any color sensation $x\in \pp$ to any other one $y\in \pp$ via a single change of background illumination would be illusory if $x$ and $y$ are very far apart in terms of chromatic attributes. The following considerations will clarify how to correctly interpret the composition of background transformations. Let us consider again Figure \ref{fig:backgroundchange} and search for a transformation $B$ such that the green sensation $x\in \pp$, $x=[(\textbf{x}_0,\textbf{b}_0)]_\sim$ of the color stimulus in the center of the image on the left is transformed into a arbitrary different color $y\in \pp$. The first transformation $B_1$ that we could be used is, for example, the one shown on the right hand side of Figure  \ref{fig:backgroundchange}. The key observation is that, thanks to what stated at the beginning of this section, the yellowish perceived color $x_1\in \pp$, $x_1=[(\textbf{x}_0,\textbf{b}_1)]_\sim$, $\textbf{b}_1 \neq \textbf{b}_0$, can be characterized by another couple $( \textbf{x}_1,\tilde{ \textbf{b}}_1)$ that matches $x_1$. Then, by performing a wisely chosen  background change $B_2$ on this alternative characterization of $x_1$, we can transform it into a color $x_2\in \pp$, $x_2=[(\textbf{x}_1,\textbf{b}_2)]_\sim$, perceptually closer to $y$ than $x_1$. As done before, $x_2$ can be characterized by another couple $(\textbf{x}_2,\tilde{ \textbf{b}}_2)$ and a third background transformation $B_3$ can be operated on this last configuration, obtaining a color $x_3\in \pp$ perceptually closer to $y$ than $x_2$. By iterating the previous steps we arrive at the match with the desired color $y$. Of course, the experimental process has to be performed painstakingly and it is likely to be very time-consuming, but the mathematical argument discussed above guarantees that the procedure can be performed within a finite number of steps. 

\begin{figure}[!ht]
	\centering
	\includegraphics[width=2.4in]{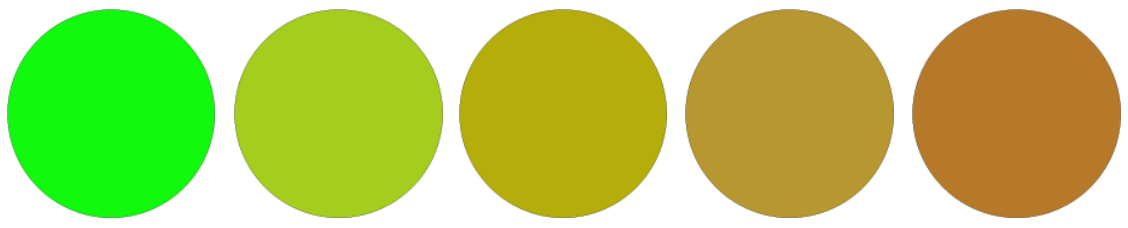}
	\caption{From left to right: the effect of composing four background transformations with the procedure explained in the text.}\label{fig:Bsteps}
\end{figure}

In Figure \ref{fig:Bsteps} we report the perceived colors $x_1,\dots,x_5$ obtained with the process described above, which shows how a color sensation can be gradually moved toward another one via composition of background transformations.

%\begin{figure}[!h]
%\begin{center}
%\includegraphics[width=5cm]{localeglobale.png}
%\end{center}
%\caption{From local to global homogeneity.}
%\end{figure}

All the considerations discussed so far justify why Resnikoff is led to postulate his own fifth axiom on the structure of the color space.

\bigskip

{\bf Axiom 5} (Resnikoff 1974, \cite{Resnikoff:74}): $\pp$ is a GL$^+(\pp)$-homogeneous space.

\bigskip
Axioms 1 to 5 imply that \textit{$\pp$ is an open convex regular homogeneous cone} (and, as such, also \textit{connected} and \textit{contractible}) embedded in a three dimensional vector space $V$. 

Such objects have been classified and this is what will allow us to explicitly determine the only possible geometrical structures of $\pp$. However, we postpone this analysis to section \ref{sec:structuresofpp} after an interlude in which we discuss an important issue related to the linearity of background transformations.

\subsection{The issue of linearity in Resnikoff's model}\label{sec:linearityissue}
The transitive action of the changes of background illuminations on $\pp$ has been extensively analyzed above. Here we concentrate on the remaining properties that they must fulfill.

Of course every $B$ preserves $\pp$ because a perceived color is still such after a background change, moreover all transformations $B$ are clearly invertible, since we can perform the reverse change and turn back to the original color sensation.

However, there is a crucial issue that Resnikoff failed to analyze: 
%the underlying linear structure of $\pp$ defined in (\ref{eq:linpp}) has been proven to be well-defined for the standard observational configuration used in colorimetry, but not for the color in (uniform) context defined by  (\ref{eq:ppincontext}). What has to be tested is the following property: if $(\textbf{x}_i,\textbf{b})$, $i=1,2$, are any two color stimuli embedded in the same background illumination and if $(\tilde{ \textbf{x}}_i, \tilde{ \textbf{b}})$, $i=1,2$, are any two different color stimuli, embedded in a different background illumination, that perceptually match the previous ones, then is it true that the couple $(\lambda_1 \textbf{x}_1+\lambda_2\textbf{x}_2,\textbf{b})$ is perceptually matched by $(\lambda_1\tilde{ \textbf{x}}_1+\lambda_2\tilde{ \textbf{x}}_2, \tilde{ \textbf{b}})$ for a reasonably extended range of scalars $\lambda_1,\lambda_2$? Up to the author's knowledge, psycho-physical results in this sense are not yet available.
\textit{it is not clear why background changes should be linear}. Actually, Resnikoff himself, in the paper \cite{Resnikoff:74bis}, published a little after \cite{Resnikoff:74}, declared this issue to be `\textit{the least verified aspect}' of the group of transformations that he considered.

\begin{figure}[!ht]
	\centering
	\includegraphics[width=3.4in]{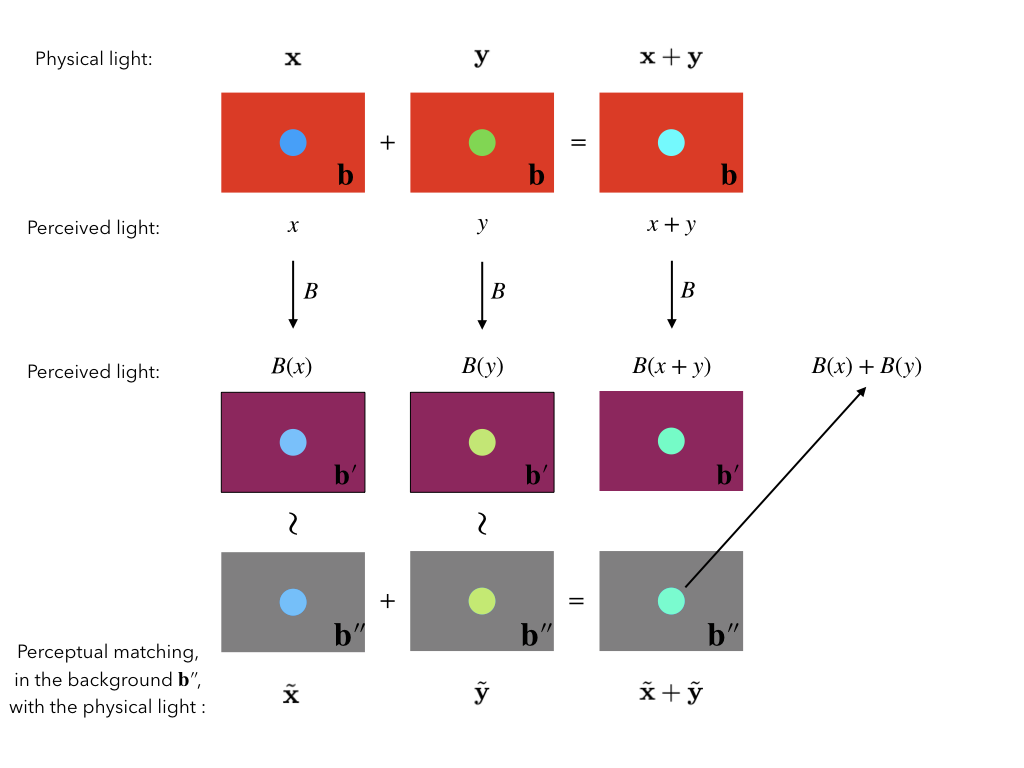}
	\caption{The experimental setup outlined in the picture can be used to check the additivity of background transformations, see the text for a detailed description.}\label{fig:Inrim_experiment}
\end{figure}

Mathematically, linear background transformations $B\in $ GL$(V)$ should behave like this on elements of $\pp$:
\beq 
B(\alpha x+\beta y)=\alpha B(x)+\beta B(y), \qquad \alpha,\beta \in \R^+, \; x,y\in \pp.
\eeq  
In Figure \ref{fig:Inrim_experiment} we outline a psycho-physical experiment to check the additivity of background transformations, a similar procedure can be used to verify if $B$ behaves linearly with respect to scaling.

Consider two physical lights $\textbf{x},\textbf{y}$ and their superposition $\textbf{x}+\textbf{y}$, all three embedded in a background $\textbf{b}$. They induce color sensations $x$, $y$ and,  assuming eq.  (\ref{eq:linpp}), $x+y$. After the change of background $B$ from $\textbf{b}$ to $\textbf{b}'$, the color sensations induced by $\textbf{x}$, $\textbf{y}$ and $\textbf{x}+\textbf{y}$ will become $B(x)=[(\textbf{x},\textbf{b}')]_\sim$, $B(y)=[(\textbf{y},\textbf{b}')]_\sim$ and $B(x+y)=[(\textbf{x}+\textbf{y},\textbf{b}')]_\sim$, respectively. 

Then consider an auxiliary background $\textbf{b}''$ and search for the physical lights $\tilde{\textbf{x}}$ and $\tilde{\textbf{y}}$ that, embedded in $\textbf{b}''$, are perceived as $B(x)$ and $B(y)$, i.e. $B(x)=[(\tilde{\textbf{x}},\textbf{b}'')]_\sim$ and $B(y)=[(\tilde{\textbf{y}},\textbf{b}'')]_\sim$, respectively. Thus, the superposition of $\tilde{\textbf{x}}$ and $\tilde{\textbf{y}}$ will give $B(x)+B(y)=[(\tilde{\textbf{x}}+\tilde{\textbf{y}},\textbf{b}'')]_\sim$. If  $B(x+y)=[(\textbf{x}+\textbf{y},\textbf{b}')]_\sim$ matches $B(x)+B(y)=[(\tilde{\textbf{x}}+\tilde{\textbf{y}},\textbf{b}'')]_\sim$, then $B$ is additive with respect to the auxiliary background $\textbf{b}''$. By repeating the test with a sufficiently diversified set of auxiliary background, the additivity of $B$ with respect to $\textbf{x}$, $\textbf{y}$ and $\textbf{b}$, $\textbf{b}'$ is tested. Finally, by varying also $\textbf{x}$, $\textbf{y}$ and $\textbf{b}$, $\textbf{b}'$, the additivity of $B$ \textit{tout court} is tested. 

Until the linearity hypothesis about the changes of background is experimentally confirmed, it remains a conjecture that lies at the core of Resnikoff's model. 

If these transformations turned out to be non-linear, this would not invalidate Resnikoff's results that we will discuss in the following section, it would just mean that the group GL$^+(\pp)$, which is supposed to act transitively on $\pp$, cannot be represented by changes of background illuminations. On the other side, the hypothesis of homogeneity of $\pp$ seems very reasonable and nothing prevents other (at the moment unknown) transformations to be identifiable with the elements of a group acting transitively on $\pp$.

\section{Consequences of the homogeneity axiom on the geometrical structure of $\pp$}\label{sec:structuresofpp}

In this section we will make use of standard results of homogeneous spaces theory to prove the most important outcome of \cite{Resnikoff:74}. Classical references are e.g. \cite{Helgason:79}, \cite{Warner:2013} and \cite{Faraut:94}.

GL$^+(\pp)$ is a Lie subgroup of GL$^+(V)$ and, for every fixed $x\in \pp$, the subgroup of GL$^+(\pp)$ defined by $K_x=\{B\in \text{GL}^+(\pp), \; B(x)=x\}$ is called the stabilizer, or isotropy subgroup, of GL$^+(\pp)$ at $x$. In terms of color perception, for a fixed perceived color $x\in \pp$, generated by a light stimulus in a given background, $K_x$ represents the subgroup of changes of background that leave the color sensation $x$ unaltered. 

The isotropy subgroups are conjugated to each other, i.e. $\forall x,y\in \pp$ $\exists \tilde B\in $ GL$^+(\pp)$ such that  $K_y=\tilde{B}K_x{\tilde{B}}^{-1}$, moreover, since $\pp$ is a homogeneous convex cone, each $K_x$ is a maximal compact subgroup of its automorphism group GL$^+(\pp)$.

The following result will be fundamental: if a differential manifold $X$ is a  $G$-homogeneous space w.r.t the action $\eta: G\times X \to X$ of a Lie group $G$ and $K_x$ is the stabilizer at $x\in X$, then the map\footnote{We recall that, given a topological group $G$, a normal subgroup $H$ of $G$ is a subgroup of $G$ such that $gH=Hg$ $\forall g\in G$, where $gH=\{gh, \; h\in H\}$ is the left coset of $H$ in $G$ w.r.t $g$ and $Hg=\{hg, \; h\in H\}$ is the right coset of $H$ in $G$ w.r.t $g$. Given a normal closed subgroup $H$ of $G$, the quotient (or factor) group $G/H$ is the group of all cosets (left or right, since they are the same because $H$ is normal) with the following group structure: $(gH) (g'H) = (gg')H$.} 
$\beta: G/K_x \to X$ defined by $\beta(gK_x)=\eta(g,x)$ is a diffeomorphism for every $x\in X$. Following Resnikoff, from now on, we will drop off the sub-index and write simply $K$.

We have all the information that we need in order to give an alternative, simpler, proof of the main result of \cite{Resnikoff:74}.

\begin{theorem}
	Axioms 1$-$5 imply that $\pp$ is diffeomorphic to either 
	\beq \label{eq:pp1}
	\pp_1 \cong \R^+ \times \R^+ \times \R^+
	\eeq 
	or
	\beq \label{eq:pp2}
	\pp_2 \cong \R^+ \times \emph{SL}(2,\R)/\emph{SO}(2).
	\eeq 
\end{theorem}
The first characterization embodies the well-known color spaces with three separated chromatic coordinates, e.g. LMS, RGB, XYZ, and so on, see e.g. \cite{Gonzales:02}. The second characterization obtained by Resnikoff is novel with respect to the classical flat color spaces and it introduces the Poincar\'e-Lobachevsky 2-D space of constant negative curvature $\text{SL}(2,\R)/\text{SO}(2)$ in color theory.

\medskip

\noindent \textbf{Proof}. Applying the results recalled above to the homogeneous color space $\pp$, we get the diffeomorphic identification: 
\beq 
\pp \cong \text{GL}^+(\pp)/K. 
\eeq
Let us rewrite every $B\in $ GL$^+(\pp)$ in the form $B=\det(B)\frac{B}{\det(B)}$, where $\det(B) \in \R^+$ and $\frac{B}{\det(B)}\in $ SL$(\pp)$, where SL$(\pp)$ is the subgroup of GL$^+(\pp)$ given by the matrices of this group with determinant 1.

It follows that GL$^+(\pp)=\R^+ \times$ SL$(\pp)$ and, since the isotropy subgroup of $\R^+$ is evidently $\{1\}$ and $\R^+/\{1\} \cong \R^+$, the only non trivial part of the quotient operation is on a closed subgroup $K$ of SL$(\pp)$, thus
\beq\label{eq:ppslpk}
\pp \cong \R^+ \times \text{SL}(\pp)/K,
\eeq 
where both $\R^+$ and $\text{SL}(\pp)/K$ are homogeneous spaces.

As differential manifolds, $\pp$ has dimension 3 and $\R^+$ has dimension 1, so $\text{SL}(\pp)/K$ has dimension 2. Plus, $\pp$ and $\R^+$ are connected and contractible, thus the expression (\ref{eq:ppslpk}) implies that also $\text{SL}(\pp)/K$ is connected and contractible. Such type of spaces have been classified by Sophus Lie \cite{Lie:1893}, see also  \cite{Komrakov:93} and \cite{Doubrov:1995} for a more modern treatise. It turns out that the only bidimensional connected contractible homogeneous spaces are either $\R^2$, diffeomorphic to $\R^+\times \R^+$ via the map $\R^2 \ni(x,y) \mapsto (\exp(x),\exp(y))\in \R^+\times \R^+$, or the hyperbolic plane SL$(2,\R) /$SO$(2)$ (see section \ref{subsec:hyperbmodel} for more details about hyperbolic spaces). \qed 

The proof provided by Resnikoff in \cite{Resnikoff:74} is not only much longer and difficult to follow, but it is also flawed. In fact, one of the fundamental arguments for his proof is the statement, at page 112, that, whatever the dimension of $\pp$, the contractility of $\cal P$ implies that the Lie group SL$(\pp)$ coincides with the exponential of its Lie algebra $\mathfrak{sl}(\pp)$. This implication, however, is not true, as we show in the Appendix with a counter-example.

\subsection{The models of the hyperbolic space SL$(2,\R)/$SO$(2)$}\label{subsec:hyperbmodel}
With the perceived color space $\pp_2$, Resnikoff introduced in colorimetry a hyperbolic space. In his 1974 paper, he acknowledges the 1962 work of H. Yilmaz \cite{Yilmaz:62} which was, historically, the first one to consider hyperbolic structures to study color perception, even though with a much less rigor than Resnikoff. 

Differently than Euclidean spaces or spheres, hyperbolic spaces can be characterized in several equivalent ways, called \textit{models}, each one useful for different purposes. For a general discussion about hyperbolic models see e.g. \cite{Martelli:2016}. Here, we just report the models of the hyperbolic space of interest for us, i.e. SL($2,\R)$/SO$(2)$:

\begin{itemize}
	\medskip
	\item the hyperboloid $I^2=\{v\in \R^3 \, : \, \langle v,v \rangle_{\cal L}=-1, \; v_3 >0  \}$, where $\langle v,v \rangle_{\cal L} = v_1^2+v_2^2-v_3^2$ is the Lorentz scalar product in $\R^3$. The equation $\langle v,v \rangle_{\cal L}=-1$ defines the two-sheet hyperboloid in $\R^3$, so that $I^2$ is the connected component with $v_3>0$, also called the upper hyperboloid sheet; 
	\medskip	
	\item the upper half plane $H=\{(x,y)\in \R^2 \, : \, y>0\}\cong\{z\in \C \, : \, \mathfrak{Im}(z)>0 \}$;
	\medskip
	\item the Poincaré disk ${\cal D}=\{(x,y)\in \R^2  \, : x^2+y^2 < 1\}\cong\{z\in \C  \, : |z| < 1\}$;
	\medskip
	\item Sym$_1^+(2,\R)$, the set of $2\times 2$ real symmetric positive-definite matrices $M$ with unitary determinant, i.e. $M^t=M$, det$(M)=1$ and $u^tMu>0$ for all $u\in \R^2$.
\end{itemize}

\medskip
This last characterization will be particularly important for the following, in fact, it will give us the possibility to interpret the elements of $\pp_2$ as matrices. To see how, let us define Sym$^+(2,\R)$ to be the set of $2\times 2$ real symmetric positive-definite matrices, any matrix $M\in $ Sym$^+(2,\R)$ can be written as $M=\det(M)\frac{M}{\det(M)}$, with $\det(M)\in \R^+$ since $M$ is positive-definite and $N=\frac{M}{\det(M)}\in $ Sym$_1^+(2,\R)$. This simple consideration implies that: 
\begin{equation}\label{eq:C2symmatrix}
\pp_2 \cong \text{Sym}^+(2,\R).
\end{equation}

\section{Selection of invariant Riemannian metrics for the color spaces ${\cal P}_1$ and ${\cal P}_2$}\label{sec:Riemannmetrics}

Once Resnikoff established the only possible geometrical structure of $\pp$ compatible with the Axioms 1-5, he searched for Riemannian metrics on $\pp$ to measure color dissimilarity. As for the geometry of $\pp$, he uniquely singled out the metrics thanks to an invariance principle.

We recall that a Riemannian metric $g$ on a differentiable manifold $X$ of dimension $n$ is a symmetric positive-definite tensor field of type $(0,2)$ on $X$, i.e. a correspondence which assigns, smoothly with respect to each point $x\in X$, a scalar product $g_x : T_xX \times T_xX \to \R$, $(v,w) \mapsto g_x(v,w)$ for all $v,w\in T_x X$, the tangent space to $X$ in $x$. A differentiable manifold $X$ endowed with a Riemannian metric $g$ is called a Riemannian manifold $(X,g)$.

Let us also recall the local coordinate expression of $g$: we fix a local chart $(U,\varphi)$ of $x$, we write with $(x^1,\ldots,x^n)$ the local coordinates of $x$ and with $(\partial_1,\ldots,\partial_n)$ the corresponding local basis of the tangent space $T_xX$.  The smooth functions $g_{\mu \nu}\in {\cal C}^\infty(U)$, $\mu,\nu=1,\ldots,n$, defined by $g_{\mu \nu}=g(\partial_\mu, \partial_\nu)$ verify $g= g_{\mu \nu} dx^\mu \otimes dx^\nu$, where Einstein's summation over repeated indices above and below is implicitly used. The component of $g_{\mu \nu}$ can be organized in a symmetric matrix and the previous expression for $g$ is often written as $ds^2 = g_{\mu \nu} dx^\mu dx^\nu$.

A Riemannian manifold $(X,g)$ is also a metric space with respect to a distance canonically induced by $g$ and defined with the help of the length of piecewise regular curves $\gamma:[0,1] \to X$. If $(X,g)$ is a connected Riemannian manifold, then, if we define the length of the curve $\gamma$ as 
\beq 
L(\gamma)=\int_0^1 \|\dot \gamma(u)\|_{\gamma(u)} \, du,
\eeq 
where $\|\dot \gamma(u)\|_{\gamma(u)}=\sqrt{g_{\gamma(u)}(\dot \gamma(u),\dot \gamma(u))}$ is the norm induced by $g$, then the function $d:X\times X\to \R^+$ defined by 
\beq 
d(x,y)=\inf\{L(\gamma), \; \gamma:[0,1]\to X \text{ piecewise regular}, \; \gamma(0)=x, \; \gamma(1)=y\} 
\eeq 
is a distance on $X$, called the Riemannian distance on $X$ induced by the Riemannian metric $g$. 

A piecewise regular curve $\gamma$ in $X$ that minimizes the Riemannian distance between a pair of points $x,y\in X$ is said to be a \textit{geodesic} connecting the two points. Thus, the Riemannian distance $d(x,y)$ can be defined as the length of a geodesic connecting $x$ to $y$.

\medskip
Let us now consider $X$ as the perceived color space $\pp$. Since Axioms 1-5 determine the geometric structure of $\pp$ as a homogeneous space, Resnikoff was naturally led to search for Riemannian metrics on $\pp$ coherent with these axioms.

If $x,y \in \pp$ are the perceived colors associated with $(\textbf{x},\textbf{b})$ and $(\textbf{y},\textbf{b})$, respectively, then, after a change of background illumination $B$ from $\textbf{b}$ to $\textbf{b}'\neq \textbf{b}$, $x$ and $y$ will be modified into $x'=B(x)\in \pp$ and $y'=B(y)\in \pp$.

\begin{figure}[!ht]
	\centering
	\includegraphics[width=3.8in]{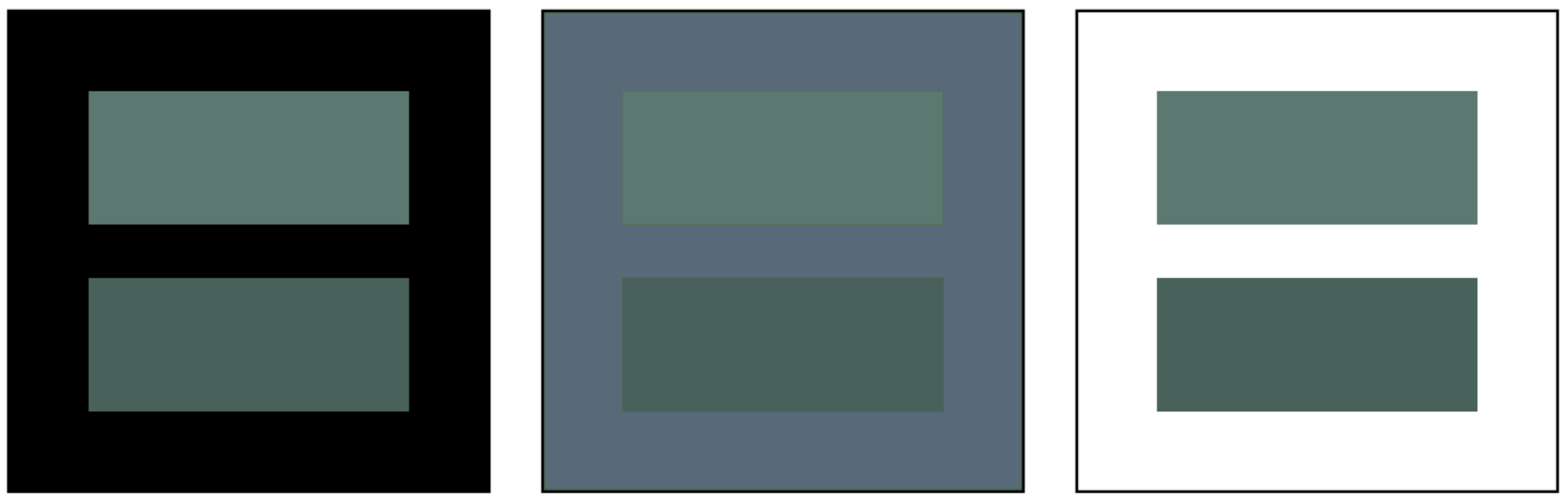}
	\caption{The crispening effect (see text) used to refute Resnikoff's Axiom 6 about the invariance of the perceptual color metric with respect to changes of background illuminations.}\label{fig:crispening}
\end{figure}

Resnikoff wanted to analyze the consequences of the following assumption (that he called Axiom 6): if $d:\pp \times \pp \to [0,+\infty)$ is the Riemannian distance on $\pp$ that measures perceptual differences between pairs of perceived colors $x,y\in \pp$, then $d$ satisfies
\beq\label{eq:dperceptual}
d(B(x),B(y))=d(x,y), \quad \forall x,y\in \pp, \; \forall B\in \text{GL}^+(\pp),
\eeq
i.e. the perceptual dissimilarity between $x$ and $y$ is the same as that between $x'$ and $y'$, or, in mathematical terms, $B$ is an isometry for the distance $d$. This assumption, however, must be refuted because it is not coherent with human color perception, as clearly shown by the \textit{crispening effect} represented in Figure \ref{fig:crispening}. The same couple of color stimuli in embedded in three different backgrounds, it is clear that \textit{the perceptual difference is not background independent}.

It is, however, very interesting to follow Resnikoff's argument and determine the metrics of $\pp$ that satisfy eq. (\ref{eq:dperceptual}), because this will point out that those metrics are not fit to represent perceptual distances for color in context.

The request of GL$^+(\pp)$-invariance permits to identify in a unique way  Riemannian metrics for $\pp_1$ and $\pp_2$. First of all, let us recall that all diffeomorphism $f:X\to Y$ induces a linear isomorphism $df_x:T_xX\to T_{f(x)}Y$, plus, if $(X, g)$ and $(Y, h)$ are Riemannian manifolds and $d_h,d_g$ are the Riemannian distances associated to the Riemannian metrics $h$ and $g$, respectively, then $f$ is an isometry, i.e. $d_h(f(x),f(y))=d_g(x,y)$ for all $x,y\in X$, if and only if:
\beq
h_{f(x)}(df_x(v),df_x(w)) = g_x(v,w), \quad \forall x\in X, \, \forall v,w\in T_xX.
\eeq
In our case, by choosing $X=Y=\pp$ and $f=B$, we have the possibility to reformulate the isometric condition expressed in eq. (\ref{eq:dperceptual}) as follows:
\beq \label{eq:pushforward}
g_{B(x)}(dB_x(v),dB_x(w))=g_x(v,w), \quad \forall B\in \text{GL}^+(\pp), \; \forall x\in \pp, \; \forall v,w\in T_x(\pp).
\eeq
Recall now that $\pp\cong$ GL$^+(\pp)/K$ and that, by homogeneity of $\pp$, for every couple of elements $x,y\in \pp$, we can write $y=B(x)$ for some $B \in$ GL$^+(\pp)$. If we select for $x$ the equivalent class to which the identity transformation of GL$^+(\pp)$ belongs, i.e. the coset $K$ itself, then, by definition, we get $B(x)=x$ for all $B\in K$. By transitivity, the $K$-invariance is independent on the choice of $x$, thus eq. (\ref{eq:pushforward}) implies that
\beq \label{eq:pushforward2}
g_{x}(dB_x(v),dB_x(w))=g_x(v,w), \quad \forall B\in K, \; \forall x\in \pp, \; \forall v,w\in T_x(\pp).
\eeq
The quest for perceptual color metrics on $\pp$ is thus reduced to the much simpler task of searching for $K$-invariant Riemannian metrics for the spaces $\pp_1$ and $\pp_2$.

\medskip
\noindent

For $\pp_1$, $K=\emptyset$, so $K$-invariance does not introduce any constraint. However, the metric must be the sum of $\R^+$-invariant metrics on each factor and all $\R^+$-invariant metrics on $\R^+$ are proportional: once we have identified one such metric, all the others are positive multiples of it. 

It is clear that a $\R^+$-invariant metric on $\R^+$ is given by $ds^2=\left(\frac{dx}{x}\right)^2$, thus the general color metric satisfying eq. (\ref{eq:dperceptual}) on $\pp_1$ is
\beq \label{eq:metric1}
ds^2= \alpha_1\left(\frac{dx_1}{x_1}\right)^2 + \alpha_2 \left(\frac{dx_2}{x_2}\right)^2 + \alpha_3 \left(\frac{dx_3}{x_3}\right)^2, \qquad \alpha_k\in \R^+, \; k=1,2,3,
\eeq
which is precisely Stiles' generalization of Helmholtz's metric (this last one corresponds to the particular case $\alpha_1=\alpha_2=\alpha_3=1$), see e.g. \cite{Wyszecky:82}.

\medskip
For $\pp_2$, $K=\text{SO}(2)$, so that the tangent space of $\pp_2$ at any $x\in \pp_2$ is
\beq 
T_x \pp_2=\R \oplus T_K \text{SL}(2,\R)/\text{SO}(2), \qquad \forall x\in \pp_2.
\eeq  
In this case, $K$-invariance means invariance under rotations, so that a background-invariant color metric for this realization of $\pp$ must be the sum of a 1-dimensional and 2-dimensional Euclidean metrics. This implies that, also for $\pp_2$, the color metric satisfying (\ref{eq:dperceptual}) is unique up to the selection of units of measure on each Cartesian factor $\R^+$ and $\text{SL}(2,\R)/\text{SO}(2)$.

An explicit characterization of such a metric on $\pp_2$ can be written by recalling from eq. (\ref{eq:C2symmatrix}) of section \ref{subsec:hyperbmodel} that $\pp_2 \cong \text{Sym}^+(2,\R)$, thus interpreting a perceived color $x$ as a $2\times 2$ positive-definite real symmetric matrix.

The action of GL$(2,\R)$ on $\pp$ is given by GL$(2,\R) \times \pp_2 \to \pp_2$, $(A,x) \mapsto AxA^t$, thus, every background transformation $B:\pp_2 \to \pp_2$ can be parameterized by a matrix $A\in $ GL$(2,\R)$ and written as follows: $B_A(x)=AxA^t$. it turns out that every GL$(2,\R)$-invariant Riemannian metric on $\pp_2$ is a scalar multiple of the so-called Rao-Siegel metric \cite{Amari:2012,Calvo:90,Siegel:14}:
\beq \label{eq:FisherRao}
ds^2 = \text{Tr}(x^{-1} dx x^{-1} dx),
\eeq 
Tr being the matrix trace. Let us verify the GL$(2,\R)$-invariance: first of all notice that $B_A(x)^{-1}={(A^t)}^{-1} x^{-1} A^{-1}$ and that, by linearity, $dB_A(x) = A \, dx \, A^t$. So 
{\small{\begin{equation}
		\begin{split}
		\text{Tr}(B_A(x)^{-1} dB_A(x) B_A(x)^{-1} dB_A(x)) & = \text{Tr}({(A^t)}^{-1} x^{-1} A^{-1} A \, dx \, A^t {(A^t)}^{-1} x^{-1} A^{-1} A \, dx \, A^t) \\
		& = \text{Tr}({(A^t)}^{-1} x^{-1}  dx  \, x^{-1}  \, dx \, A^t),
		\end{split}
		\end{equation}}}
by using the cyclic property of the trace we have
\begin{equation}
\text{Tr}(B_A(x)^{-1} dB_A(x) B_A(x)^{-1} dB_A(x)) = \text{Tr}(x^{-1}  dx  \, x^{-1}  \, dx),
\end{equation}
$\forall B_A\in $ GL$(2,\R)$, thus confirming the GL$(2,\R)$-invariance. 

To resume, \textit{the Helmholtz-Stiles metric on $\pp_1$ and the Rao-Fisher metric on $\pp_2$ cannot be considered perceptual metrics for color in context} since they violate  the crispening effect. For color in context we mean color stimuli perceived in a visual scene in which the background can undergo temporal and/or spatial changes. Of course, the crispening effect does not disqualify the metrics above when the background is fixed, however, in this case, we cannot use anymore the argument about the invariance under background changes to single them out.

In 1974, at the time of Resnikoff's paper \cite{Resnikoff:74}, only a few papers about this perceptual phenomenon were available, this explains why, not being aware of it, he wrongly assumed that a perceptual color metric should be background-invariant.

\section{Conclusions}

A detailed analysis of the Resnikoff model and his homogeneity axiom for the space of perceived colors $\pp$ led us to the following results: first of all we have provided an alternative and much simpler proof of the main result contained in \cite{Resnikoff:74}, i.e. the existence of only two geometric structures compatibles with Schrödinger's axioms together with Resnikoff's homogeneity one. We have also shown, via a counter-example, that the original proof is flawed by a mathematical assumption that is not true.

Secondly, we have underlined the exigence of developing psycho-physical experiments to check the linearity of background transformations that Resnikoff supposed to be identified with the group of symmetries acting transitively on $\pp$. A proposal for such an experiment has been detailed.

Finally, we have discussed Resnikoff's hypothesis about the isometric character of background transformations with respect to the Riemannian metrics on $\pp$ that should represent perceptual color differences in his theory. The crispening effect shows that Resnikoff's hypothesis must be refuted, thus, both the Helmholtz-Stiles and the Fisher-Rao metrics, singled out by using this hypothesis, cannot be perceptual color distances in context.

In the second half of this two-part paper, $\pp$ will be analyzed by using a different strategy that relies on Jordan algebras. The link between the two parts is given by the following consideration: Schr\"odinger's axioms imply that $\pp$ is a regular convex cone embedded in a real vector space of dimension three, if we accept  Resnikoff's homogeneity axiom, then $\pp$ becomes an open regular homogeneous convex cone. By adding a last hypothesis, the so-called self-duality, $\pp$ becomes a \textit{symmetric cone} and it turns out that these objects can be identified with the positive elements of a (formally real) Jordan algebra. The rich mathematical results associated to Jordan algebras will permit to build a novel, quantum-like, theory of $\pp$.

\section{Appendix}\label{sec:appendix}
We discuss in this appendix a counter-example which shows that it is not true, as claimed by Resnikoff, that the contractility of $\pp$ implies $\exp(\mathfrak{sl}(\pp))=$ SL$(\pp)$.

The argument can be discussed already for the case SL$(\pp)=$ SL$(2,\R)$ and its Lie algebra $\mathfrak{sl}(2,\R)$, given by the real $2\times 2$ traceless matrices. If the exponential map $\exp:\mathfrak{sl}(2,\R) \to $ SL$(2,\R)$ were onto, then the matrix
$$ T=\begin{pmatrix}
-1 & 1 \\
0 & -1
\end{pmatrix} \in \text{SL}(2,\R)$$
could be written as $T=\exp(A)$, for a suitable $A\in \mathfrak{sl}(2,\R)$. Thanks to the well-known Schur decomposition theorem, $A$ is similar to an upper triangular matrix $U$ whose diagonal elements are the eigenvalues of $A$, i.e. $A=PUP^{-1}$, where $P\in $ GL$(2,\C)$. However, thanks to its cyclic property, the trace is similarity-invariant, so  Tr$(A)=$ Tr$(U)$ and, being Tr$(A)=0$, it follows that the $U$ must have the following form
$$ U=\begin{pmatrix}
\alpha & \beta \\
0 & -\alpha
\end{pmatrix},$$
$\alpha,\beta\in \C$, so that 
$$A=P\begin{pmatrix}
\alpha & \beta \\
0 & -\alpha
\end{pmatrix}P^{-1},$$
$\alpha$ and $-\alpha$ being the eigenvalues of $A$. Recalling that $\exp(A)=\sum\limits_{n\in \N} \frac{A^n}{n!}$, we have that
$$ T=\exp(A)=\sum_{n\in \N}\frac{(PUP^{-1})^n}{n!}=P\left(\sum_{n\in \N}\frac{U^n}{n!}\right)P^{-1} =P\exp \begin{pmatrix}
\alpha & \beta \\
0 & -\alpha
\end{pmatrix} P^{-1}.$$ 
We can now show the contradiction. First of all, if $\alpha \neq 0$, then the Schur decomposition theorem guarantees that $\alpha$ and $-\alpha$ are two distinct eigenvalues of the $2\times 2$ matrix $A$, i.e. $A$ is similar to a diagonal matrix: it exists $Q\in $ GL$(2,\C)$ such that $A=QDQ^{-1}$, with $D=$ diag$(\alpha,-\alpha)$. But then, $T=\exp(A)=Q\exp(D)Q^{-1}$, with $\exp(D)=$ diag$(e^\alpha,e^{-\alpha})$, which contradicts the fact that $T$ is clearly not diagonalizable.  

If, instead, $\alpha=0$, then
$$\exp \begin{pmatrix}
0 & \beta \\
0 & 0
\end{pmatrix} = \begin{pmatrix}
1 & 0 \\
0 & 1
\end{pmatrix} + \begin{pmatrix}
0 & \beta \\
0 & 0
\end{pmatrix} + \sum_{n=2}^\infty \frac{1}{n!} \begin{pmatrix}
0 & 0 \\
0 & 0
\end{pmatrix} = \begin{pmatrix}
1 & \beta \\
0 & 1
\end{pmatrix},$$
which implies Tr$(T)=-2\neq$ Tr$\left(\left.\exp(U)\right|_{\alpha=0}\right)=2$, but this  cannot be true because it contradicts the similarity-invariance of the trace.

\section*{Acknowledgements}
This paper is part of a program for a geometric re-foundation of colorimetry inspired by the brilliant work of H.L. Resnikoff (1937-2018). This paper is dedicated to his memory. The counter-example and the novel proof of the main result of section 4 have been developed during the collaboration with Francesco Bottacin for the supervision of the master thesis of Fiammetta Cirrone at the University of Padova, both of them are warmly acknowledged.

\bibliographystyle{plain} % Style BST file (bmc-mathphys, vancouver, spbasic).
\bibliography{bibliography}   

\end{document}